\documentclass[superscriptaddress,twocolumn, amsmath]{revtex4}
\usepackage{graphicx}
\usepackage{amsmath}
\usepackage{amsfonts}
\usepackage{amssymb}
\usepackage{placeins}
\usepackage{color}
\usepackage{hyperref}

\begin{document}

\title{Repulsion in controversial debate drives public opinion into fifty-fifty stalemate}
\author{Sebastian M.\ Krause}
\affiliation{University of Duisburg-Essen, Duisburg, Germany}
\affiliation{University of Bremen, Bremen, Germany}
\affiliation{Institute Rudjer Boskovic, Zagreb, Croatia}
\author{Fritz Weyhausen-Brinkmann}
\affiliation{University of Bremen, Bremen, Germany}
\author{Stefan Bornholdt}
\email{bornholdt@itp.uni-bremen.de}
\affiliation{University of Bremen, Bremen, Germany}
\affiliation{Santa Fe Institute, 1399 Hyde Park Road, Santa Fe, New Mexico 87501, USA}
\begin{abstract}
Opinion formation is a process with strong implications for public policy. In controversial debates with large consequences, the public opinion is often trapped in a fifty-fifty stalemate, jeopardizing broadly accepted political decisions. Emergent effects from millions of private discussions make it hard to understand or influence this kind of opinion dynamics. 
Here we demonstrate that repulsion from opinions favors fifty-fifty stalemates. We study a voter model where agents can have two opinions or an undecided state in-between. In pairwise discussions, undecided agents can be convinced or repelled from the opinion expressed by another agent. If repulsion happens in at least one of four cases, as in controversial debates, the frequencies of both opinions equalize. Further we include transitions of decided agents to the undecided state. If that happens often, the share of undecided agents becomes large, as can be measured with the share of undecided answers in polls. 
\end{abstract}
\maketitle

\section{Introduction}

Opinion formation is driven by big influencers as media~\cite{page1987moves}, social networks~\cite{barbera2015tweeting,del2016spreading} and private discussions~\cite{castellano2009statistical}. One of the most influential models for private discussions in opinion formation is the stochastic agent based voter model \cite {clifford_model_1973, holley_ergodic_1975, castellano2009statistical, redner2019review}. Its success is due to the following reasons: First, the basic idea of the model is simple and intuitive, with binary opinions (e.g.\ ``yes'' or ``no''), and pair interactions to represent discussions with a possible convincing outcome. Second, the emergent dynamics of the model is interesting. Studied on the lattice with interactions among neighbors, it is one of the few analytically solvable non-equilibrium models \cite{redner2001guide}. In two dimensions, it shows a special coarsening without surface tension, and a special kind of criticality in the presence of absorbing states \cite{dornic_critical_2001, krause_mean-field-like_2012}. Third, variations of the model have been proven interesting and realistic enough to be compared with real world data: A new kind of phase transition in an adaptive network model has been found which is consistent with the size distribution of religions \cite{holme2006nonequilibrium}. In a stock market model with a global feedback, bursty dynamics results in broadly distributed price jumps \cite{krause_opinion_2012}. A model including tactical voting was found to explain close results of the two leading candidates in elections \cite{araujo2010tactical}, and a model including mobility patterns was able to predict long range correlations in US presidential elections \cite{fernandez2014voter}. 
Some studies analyze how an intermediate state between the two opposite opinions alters the dynamics. An additional centrist opinion (as for example a central political party) was considered \cite{vazquez2004ultimate, vazquez2003constrained}, and a third state representing the coexistence of both opinions was used for language dynamics \cite{castello2006ordering,castello2009consensus}. Intermediate states were also considered as undecided states \cite{dall2007effective, dall2008algebraic, malarz2009indifferents, Svenkeson2015reaching}. 

Often the public opinion about a controversial topic is trapped around a fifty-fifty stalemate as in the Swiss referendum against mass immigration of 2014, in the 
British decision to leave the European Union of 2016 \cite{vasilopoulou2016uk} and in the presidential election of the USA in 2016. In such cases the debate can be dominated by provocative and controversial statements of prominent politicians, thus increasing the emotional and controversial character of public debate. This seems to drive the public opinion into fifty-fifty stalemates with very close results of referendums or elections. To our knowledge it is unclear which mechanism is responsible for this result. Implications are enormous, with highly polarized, split societies having two groups of almost identical size, where each group hopes for overtaking the political power with repeated referendums or early elections. 
\\

Here we identify a mechanism driving the public opinion into fifty-fifty stalemate which is typical for controversial debates. We define a simple generalization of the voter model including an undecided state between the two opinions. We consider pairwise discussions on the complete interaction matrix. Besides the possibility that an agent can be convinced by an opinion, we consider two other intuitive outcomes of a discussion: An opinion can repel (similar to contrarian behavior \cite{galam2004contrarian}), and an agent can doubt its former opinion and switch to the undecided state. We find that a clear majority opinion develops if an undecided agent is much more likely convinced than repelled by an opinion. On the other hand, if an agent is repelled in at least one of four cases instead of being convinced, as in controversial discussions, neither of the opinions is able to win in the long run. If agents often doubt their opinion the share of undecided agents becomes large. Comparing with poll data where an answer ``don't know'' is possible we find that undecided agents in our model are more easily convinced than decided agents start to doubt.
How strong agents doubt has no influence on whether a clear majority forms.

\section{Model description}\label{sec:model}

\begin{figure}[htb]
\begin{center}
	\includegraphics[width=1.0\columnwidth]{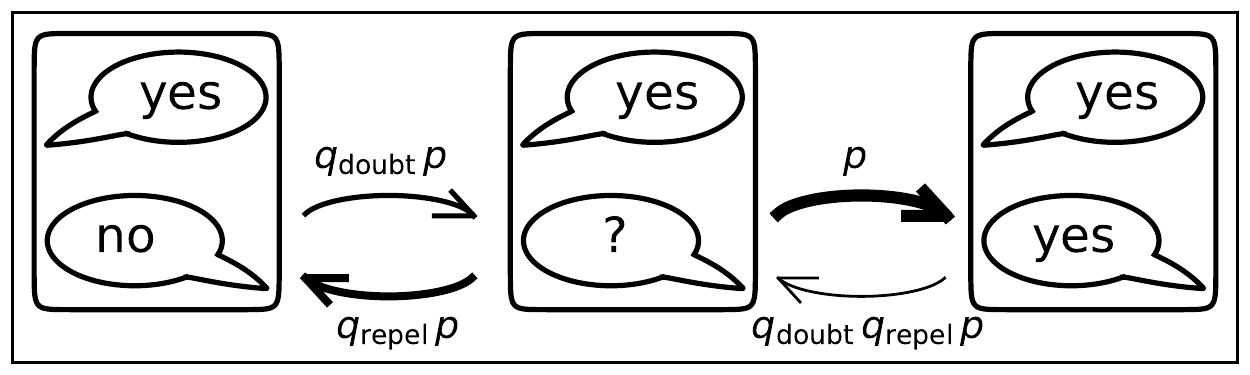}
	\caption{Sketch of the model dynamics with an intermediate state ``0'' 
	and two opinions ``+1'' and ``-1'', here named as ``?'', ``yes'' and ``no''. 
	In pairwise interactions one agent is the speaker, here the upper agent with opinion ``yes''.  
        The other agent only adopts to defined opinions ``yes'' or ``no'' of the speaker.
	Shown are transition probabilities for the second agent 
	(the scheme is identical for inverted signs). The probability $p$ for 
	convincing an agent fixes the time scale, the parameters $q_{\rm repel}$ and $q_{\rm doubt}$ tune the effects of repulsion and doubt.}
	\label{fig:model}
\end{center}
\end{figure}

We have a population of $N$ agents, where an agent $i\in \{1,\dots,N\}$ has an opinion 
$o_i\in \{-1,0,1\}$. This state is meant to describe an opinion on a 
question with two possible answers, here reduced for a clearer 
understanding to ``yes'' ($o_i=1$) and ``no'' ($o_i=-1$). 
The third possible state $o_i=0$ is meant to describe an 
undecided agent. 

The dynamics of our model is caused by changes of opinions from or to 
undecided states in the presence of decided agents. More precisely, 
in every time step we randomly choose an agent $j$ who is the speaker 
which means he puts forward an argument in favor of his opinion. 
A second agent $i$ considers this argument and  
might change his opinion. We want to describe the outcome 
of the conversation in a probabilistic way. This is illustrated in 
Figure~\ref{fig:model}. On the outer left, two agents 
with contrary opinions discuss. In the middle, one of the agents is 
undecided. On the outer right, both agents agree. 
We have to consider the following cases of speaker's and second agent's 
opinions: 

\begin{itemize}
\item If the speaker $j$ has undecided opinion $o_j=0$, nothing happens, as we assume 
that an undecided agent has no effect on other agents. 

\item If the speaking agent has decided opinion $o_j\neq 0$ and the second 
agent is undecided with $o_i=0$, the updated opinion is
$o_i'=o_j$ with probability $0<p<1$. This is the case, 
where $j$ convinces $i$. With probability $0\leq p q_{\rm repel}<1$, the 
updated opinion is $o_i'=-o_j$. An undecided agent $i$ may be faced with 
arguments used by agent $j$ which actually convince him of 
the opposite. 
With probability of $1-p- p q_{\rm repel}$ the opinion 
does not change, $o_i'=o_i$. The model parameters $p$ and $q_{\rm repel}$ 
have to be set such that this probability is positive. 

\item If speaker $j$ and second agent $i$ have opposite decided opinions 
$o_i=-o_j\neq 0$, 
$i$ changes its state to $o_i'=0$ with probability $0\leq p q_{\rm doubt}<1$. 
That is the case where agent $i$ questions its opinion when confronted 
with arguments of a speaker with opposite opinion. With 
$1-p q_{\rm doubt}$ the opinion stays $o_i'=o_i$. 

\item If speaker $j$ and second agent $i$ have the same decided opinion, $o_i=o_j\neq 0$, 
$i$ changes its state to $o_i'=0$ with probability 
$p q_{\rm doubt} q_{\rm repel}$. That is the 
case where agent $i$ is confronted with arguments of someone with 
the same opinion, with the effect that he starts to doubt. 
With $1-p q_{\rm doubt} q_{\rm repel}$ the opinion stays $o_i'=o_i$. 
\end{itemize}

For setting the time scale, we combine $N$ steps to an increase of 
one time unit, $t\to t+1$. In this way, every agent has on 
average the chance for one update per time step, independent of the total number 
of agents $N$. To characterize the macroscopic state of the system, 
we use the following macroscopic variables: The frequencies  
$\pi_0$, $\pi_1$, and $\pi_{-1}$ describing the frequencies of 
undecided agents and, respectively, of agents with opinion ``1'' or ``-1''. 
Alternatively, we use the majority opinion indicator $m=\pi_1-\pi_{-1}$. 
Using only $u=\pi_0$ and $m$, we account for the fact that 
$\pi_0$, $\pi_1$ and $\pi_{-1}$ are dependent, as they sum up to one.

\section{Model dynamics compared to poll results}\label{sec:dynamics}

\begin{figure}[htb]
\begin{center}
	\hspace{5mm}\includegraphics[width=0.855\columnwidth]{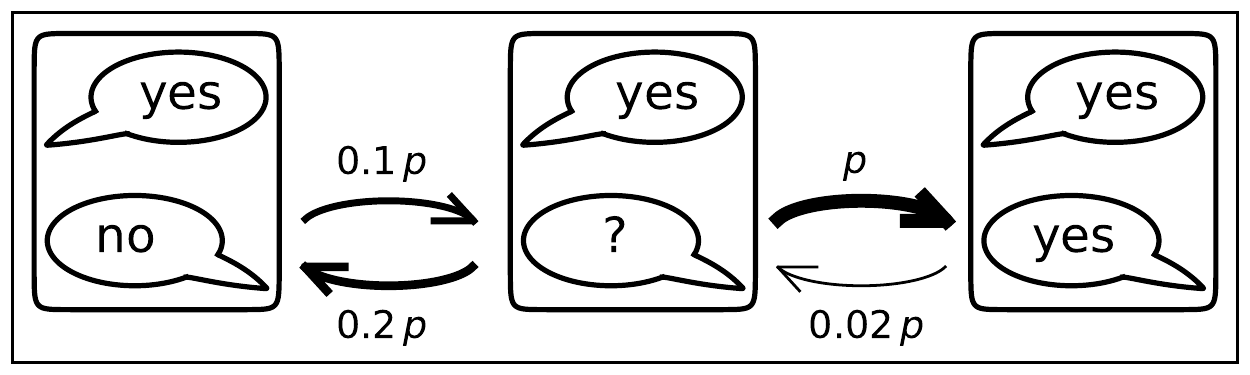}
	\includegraphics[width=1.0\columnwidth]{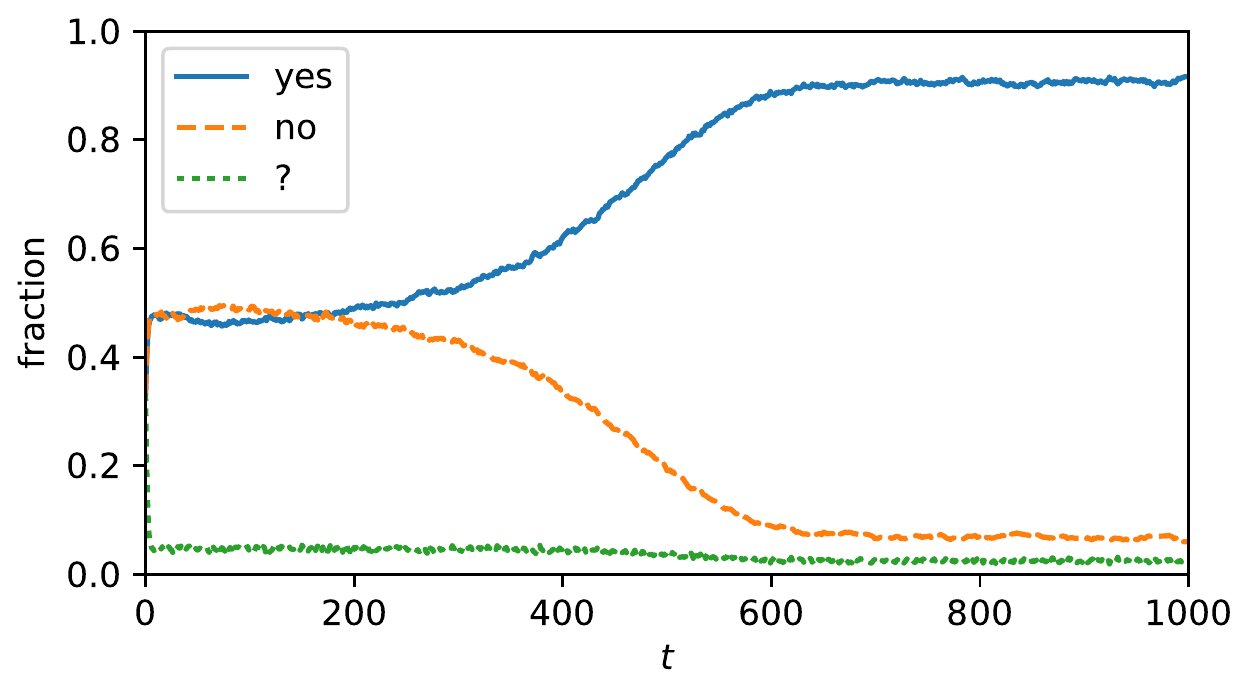}
	\caption{Time dependent opinion fractions for small repulsion parameter $q_{\rm repel}=0.2$ and small $q_{\rm doubt}=0.1$.}
	\label{fig:trajectory1}
\end{center}
\end{figure}

In Figure~\ref{fig:trajectory1} results of a simulation are shown for non-controversial debate with small repulsion parameter $q_{\rm repel}=0.2$. Here it is five times more likely for an undecided agent to be convinced than being repelled. Also the doubting parameter $q_{\rm doubt}=0.1$ is small. The number of agents is $N=5000$, the probability to convince an undecided agent is $p=0.5$. From an initial configuration with $\pi_0=\pi_1=\pi_{-1}=1/3$ the share of undecided agents (green dotted line) drops within a few time steps, the decided shares (blue solid line for ``yes'' and orange dashed line for ``no'') increase accordingly. In real populations such a fast initial dynamics is most likely not observable, because it is unrealistic that a debate is completely absent in a society and then starts abruptly. What is observable are sudden changes due to suddenly changing parameters, as we will see below. 
The shares of decided opinions stay close to each other with switching majority in the first 200 time steps. This is a reminder of the symmetry of the model. Later the shares of decided opinions slowly diverge from each other and reach a stable fixed point, while the share of undecided agents decreases slowly. We find a time scale separation between the dynamics of the share of undecided agents $u$ falling from one third to about 5\% within a few time units and the majority opinion indicator $m$ which needs hundreds of time units to change. All shares reach a steady state after $t=600$. 

\begin{figure}[htb]
\begin{center}
	\hspace{5mm}\includegraphics[width=0.855\columnwidth]{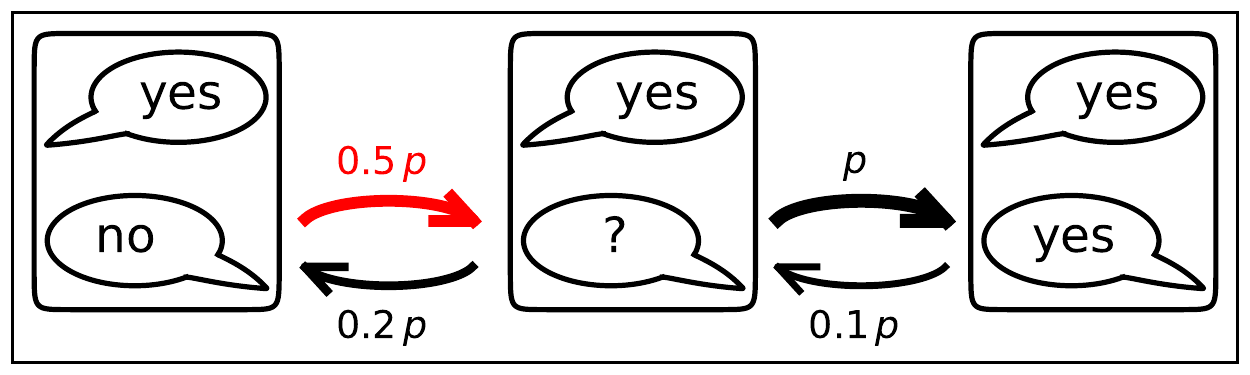}
	\includegraphics[width=1.0\columnwidth]{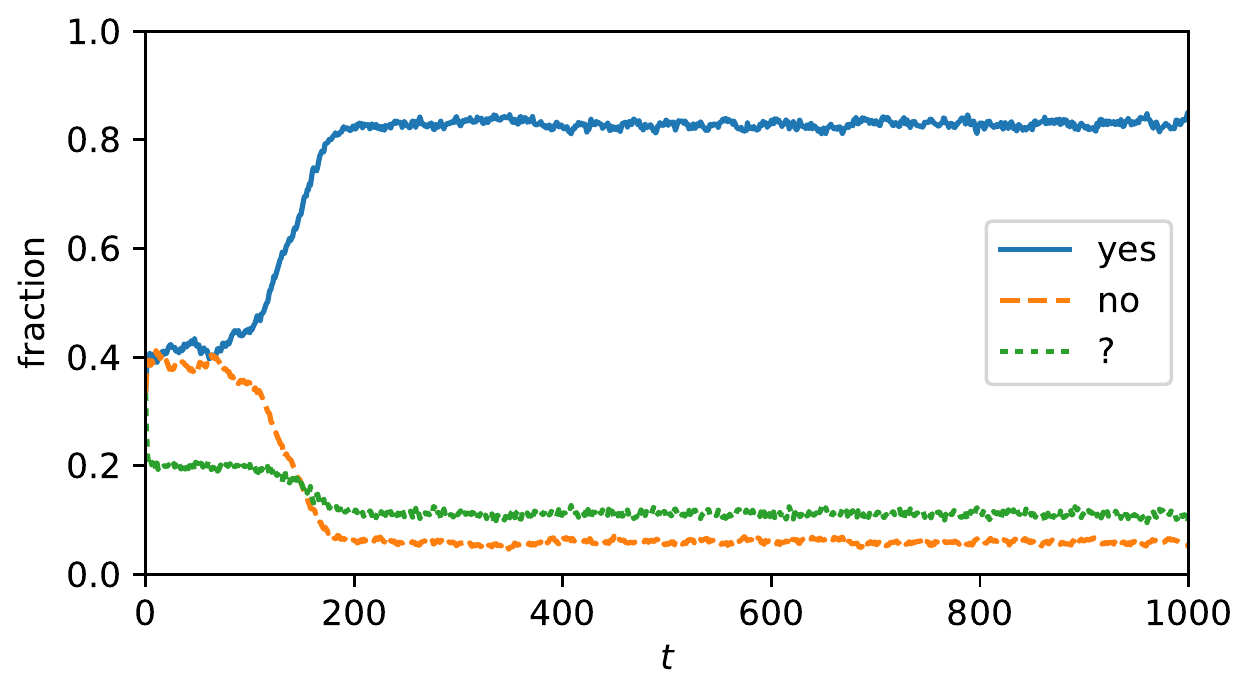}
	\caption{Time dependent opinion fractions for small repulsion parameter $q_{\rm repel}=0.2$ and large $q_{\rm doubt}=0.5$.}
	\label{fig:trajectory2}
\end{center}
\end{figure}

Figure~\ref{fig:trajectory2} reports the fractions of opinions in a simulation with strongly doubting agents with $q_{\rm doubt}=0.5$. The dynamics is similar to the case discussed before. The largest differences are a larger share of doubting agents in the steady state which is also reached much faster. As opinion changes are only possible through undecided agents, a large share of undecided agents speeds up the opinion formation. On the other hand, opinion formation is particularly slow if agents are only rarely doubting. 

\begin{figure}[htb]
\begin{center}
	\includegraphics[width=1.0\columnwidth]{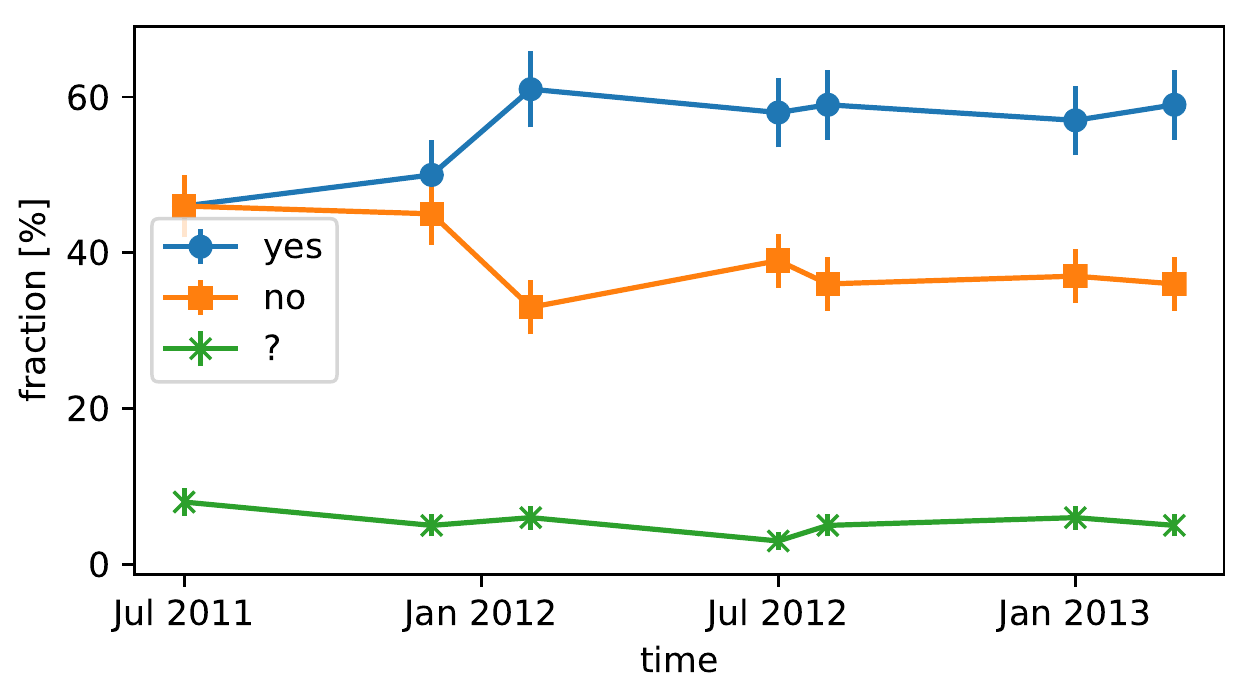}
	\caption{Results of a repeated poll performed by ``Infratest Dimap''~\cite{infratest} about the satisfaction of Germans with chancellor Angela Merkel's management of the EU financial crisis. }
	\label{fig:poll2}
\end{center}
\end{figure}

Results of a repeated poll in Fig.~\ref{fig:poll2} illustrate that the share of undecided answers (green crosses, connecting lines are to guide the eye) can indeed stay at a large level with only little variation over long times, while the shares of decided opinions perform a clear movement at the beginning and later stabilize with a clear majority of ``yes'' (blue circles).

\begin{figure}[htb]
\begin{center}
	\hspace{5mm}\includegraphics[width=0.855\columnwidth]{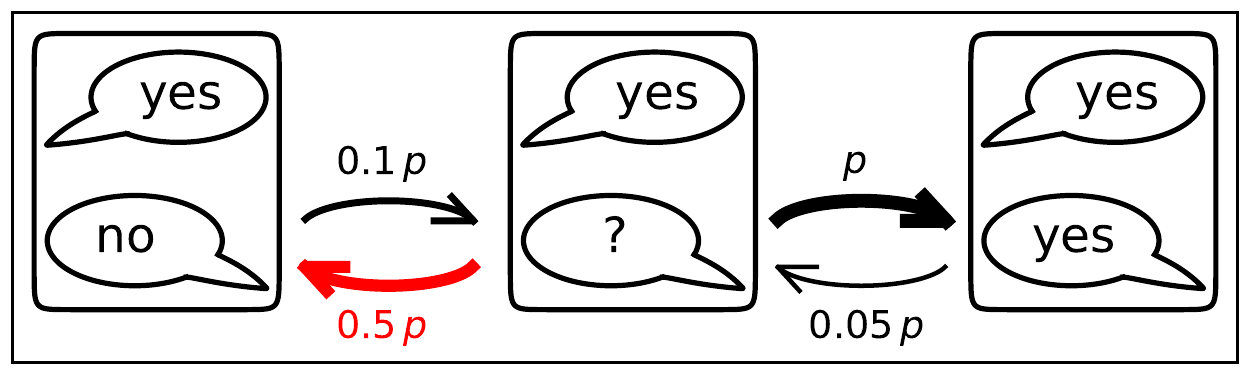}
	\includegraphics[width=1.0\columnwidth]{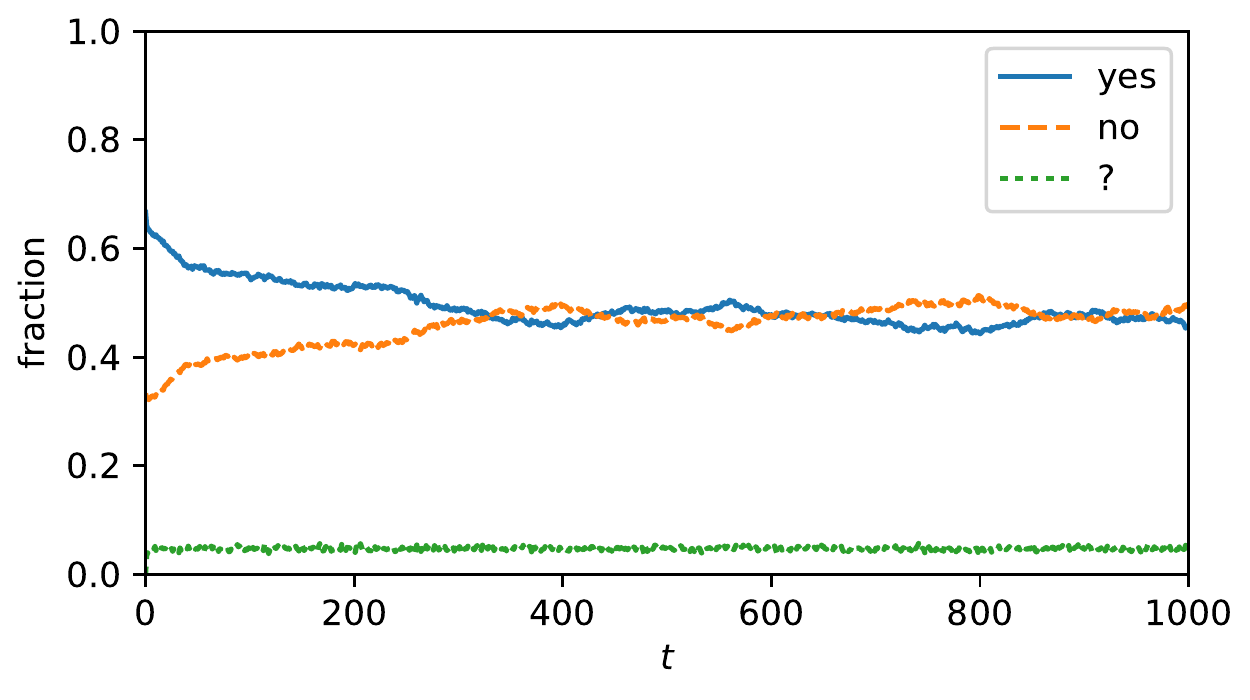}
	\caption{Time dependent opinion fractions for a controversial debate with large repulsion parameter $q_{\rm repel}=0.5$ and small $q_{\rm doubt}=0.1$.}
	\label{fig:trajectory3}
\end{center}
\end{figure}

Let us finally discuss a controversial debate with Fig.~\ref{fig:trajectory3}. The simulation here has a large repulsion parameter of $q_{\rm repel}=0.5$ such that an undecided agent is only twice more likely convinced than being repelled. The doubt parameter is returned to a small value $q_{\rm doubt}=0.1$. The initial condition is chosen asymmetric in order to clarify the convergence to the steady state, with $\pi_0=0$, $\pi_1=2/3$ and $\pi_{-1}=1/3$. After the system reaches the steady state, both decided opinion shares fluctuate around the same value. Fluctuations are so large and slow that for some time periods a clear majority opinion seems to exist, having about 10\% more than the minority opinion. 

\begin{figure}[htb]
\begin{center}
	\includegraphics[width=1.0\columnwidth]{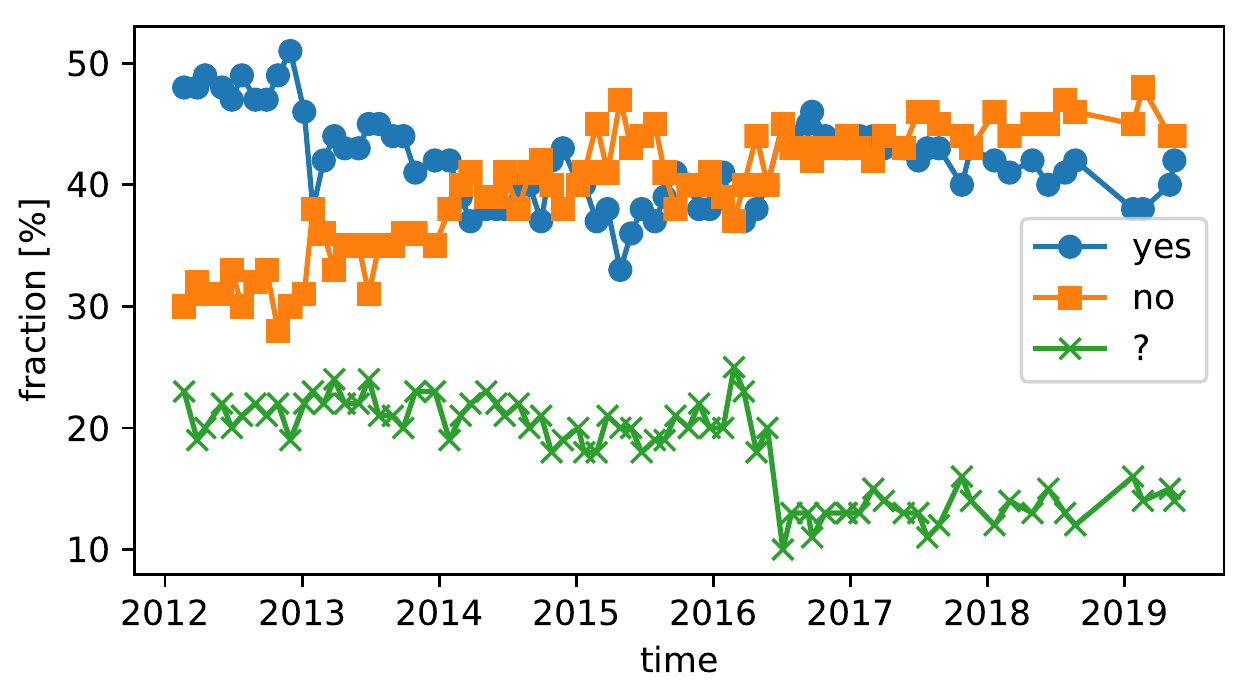}
	\caption{Results of a repeated poll performed by ``YouGov'' and accessed through ``whatukthinks.org''~\cite{brexit} on how British people would vote on a referendum to leave the EU.}
	\label{fig:poll2}
\end{center}
\end{figure}

Results of a repeated poll asking whether the UK should leave the EU (the so called Brexit) are shown in Fig.~\ref{fig:poll1}. Beginning with 2014 these data illustrate that opinions can be trapped close to each other for many years. For some periods a clear majority seems to exist, but then the majority opinion changes again. This is qualitatively the same as in our model results in Fig.~\ref{fig:trajectory3}. The share of undecided answers to the Brexit poll, here combined answers ``Don't know'' and ``Would not vote'', stay on large values until mid of 2016, where the actual Brexit referendum takes place on 23 June 2016. Then they jump to smaller values. Between the polls on 24 May 2016 and on 4 July 2016 the undecided opinion share drops to half from 20\% to 10\%. 
Within the framework of our model this can be interpreted as follows: Before the Brexit referendum many British people were rather indifferent towards this topic. This can be modeled with large $q_{\rm doubt}$ because this also can indicate a loss of decided opinion due to a loss of interest in the question. After the referendum took place, a debate about consequences of the Brexit started developing which made people realize how they are actually really affected by the Brexit. This would be modeled with smaller $q_{\rm doubt}$ and decrease the share of undecided agents as we saw before. The time scale separation as found for our model is also found in the poll data. 
The sudden jump of the undecided opinion share within a few weeks happens much faster than the slow moves of decided opinion shares on the time scale of years.

\section{Fixed point analysis}\label{sec:fp}

In all model trajectories of opinion shares discussed so far we saw convergence to a fluctuating steady state. This is connected with stable fixed points of the drift present in our model. These fixed points are calculated in the following. At fixed points the drift of majority opinion indicator $m$ and share of undecided agents $u$ vanishes. For calculating the drift we use probabilities for changes of $m$ and $u$ due to opinion changes of a single agent during a time step $\Delta t=1/N$. The probability $p_m^+$ for $m$ increases to $m+1/N$ is 
\begin{align}
p_m^+ &= \pi_0\pi_1 p  +  \pi_0\pi_{-1} p q_{\rm repel}  +
\pi_{-1}\pi_1 p q_{\rm doubt}  \nonumber\\
 &\qquad + \pi_{-1}\pi_{-1} p q_{\rm doubt} q_{\rm repel}, 
\end{align}
if we allow for the small probability event of choosing the same agent 
twice which then would have to discuss with himself. The probability $p_m^-$ for decreasing $m$ looks similar, starting from $p_m^+$ the signs of $\pi_1$ and $\pi_{-1}$ have to be switched. The drift is calculated as
\begin{align}
\dot{m}_{\rm drift}=\lim_{N\to \infty}\frac{\Delta m}{\Delta t}(p_m^+-p_m^-). 
\end{align}
We further replace with $m=\pi_1-\pi_{-1}$, $1-u=\pi_1-\pi_{-1}$ and use $\Delta m=1/N$. The procedure for calculating the drift of $u$ is the same. The result is 
\begin{align}
\frac{\dot{m}_{\rm drift}}{p} &=  m (1- q_{\rm repel})\nonumber\\
&\, \quad -m (1-u) \left[1-q_{\rm repel}(1-q_{\rm doubt})\right],\\
\frac{\dot{u}_{\rm drift}}{p} &= -m^2 \frac{q_{\rm doubt}}{2} (1-q_{\rm repel}) -(1-u) (1+q_{\rm repel})\nonumber\\
&\,\quad + (1-u)^2 (1+q_{\rm repel})\left(1+\frac{q_{\rm doubt}}{2}\right) .\label{eq:u_dot}
\end{align}
The convincing probability $p$ only fixes the time scale. 

\begin{figure}[htb]
\begin{center}
	\includegraphics[trim=33 15 13 0,clip,width=0.49\columnwidth]{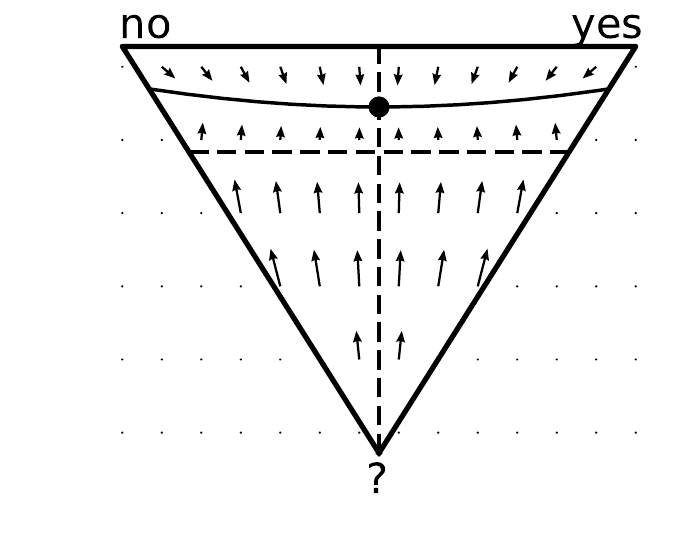}
	\includegraphics[trim=33 15 13 0,clip,width=0.49\columnwidth]{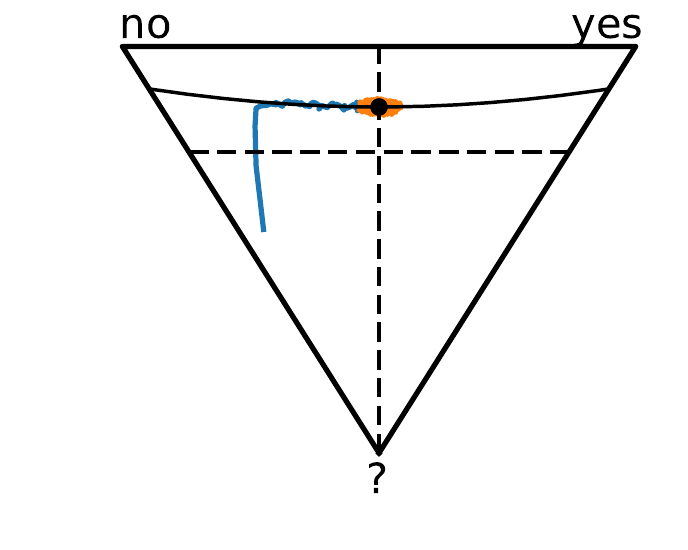}
	\caption{Left: Drift velocity field (arrows) depending on opinion shares for controversial debate with $q_{\rm repel}=0.5$ and $q_{\rm doubt}=0.35$. Line of vanishing vertical drift velocity ($\dot{u}_{\rm drift}=0$, solid line) and lines of vanishing horizontal drift velocity ($\dot{m}_{\rm drift}=0$, dashed lines). The circle indicates the stable fixed point. Right: Trajectory of opinion shares converging to steady state (blue line) and in the steady state (orange). }
	\label{fig:triangle1}
\end{center}
\end{figure}

In Fig.~\ref{fig:triangle1} opinion shares are presented in triangles. In the upper right corner the share of ``yes'' is one, on the opposing edge it is zero. The majority opinion indicator $m$ is negative on the left of the triangle and positive on the right, $u$ is zero on the top edge and one on the bottom corner. On the left of the figure the field of drift velocity points primarily away from very small or large shares of undecided opinions. In-between, there is a shaped line of vanishing drift velocity $\dot{u}_{\rm drift}=0$, indicated with the black solid line. Symmetric between the decided opinions there is a line of vanishing horizontal drift velocity $\dot{m}_{\rm drift}=0$ (vertical dashed line). On the crossing of $\dot{u}_{\rm drift}=0$ and $\dot{m}_{\rm drift}=0$, both components of drift velocity are zero and a fixed point sits on this place. This fixed point is stable because all flows point towards it. On the right of the Figure an opinion trajectory is shown on the way to the steady state (blue line). First it moves vertically in a short time until $t=4$ to the line of vanishing $\dot{u}_{\rm drift}$, then it follows this line slowly heading to the fixed point. We have a clear separation of time scales for the dynamics of $u$ and $m$, as discussed before. In the steady state (orange) the trajectory fluctuates around the stable fixed point. 

\begin{figure}[htb]
\begin{center}
	\includegraphics[trim=33 15 13 0,clip,width=0.49\columnwidth]{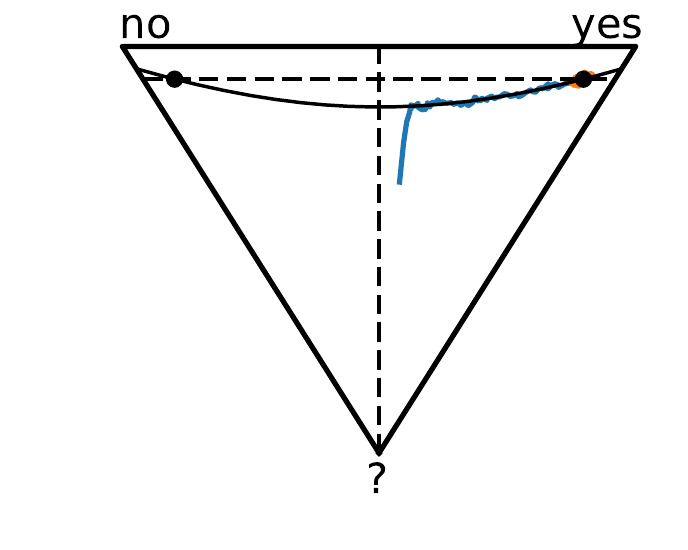}
	\includegraphics[trim=33 15 13 0,clip,width=0.49\columnwidth]{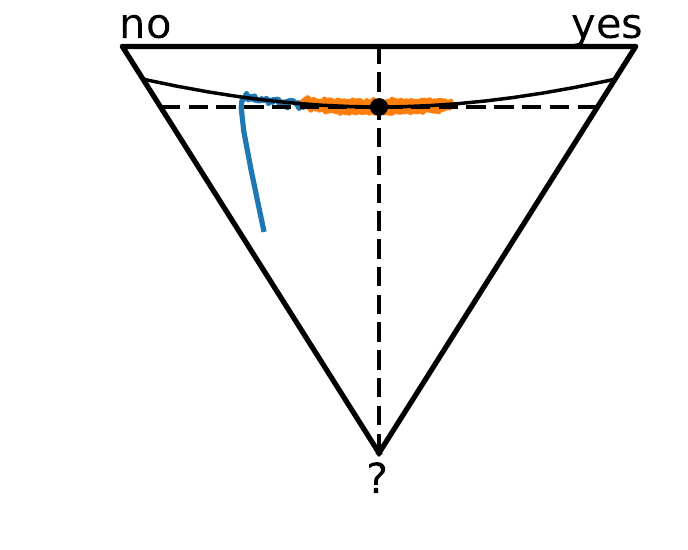}
	\caption{Left: Opinion trajectory as in Fig.~\ref{fig:triangle1} here in the non-controversial regime with $q_{\rm repel}=0.2$. The circles indicate two stable fixed points with clear majority. Right: Opinion trajectory at the phase transition with $q_{\rm repel}=1/3$.}
	\label{fig:triangle2}
\end{center}
\end{figure}

In Fig.~\ref{fig:triangle2} on the left we see how small $q_{\rm repel}$ leads to steady states with clear majority opinion. Here the line of vanishing velocity $\dot{u}_{\rm drift}=0$ and the horizontal line with $\dot{m}_{\rm drift}=0$ cross. Additionally an unstable fixed point exists in the center. The transition of one stable fixed point losing stability and at the same time two stable fixed points emerging is also called pitchfork bifurcation. On the right we see what happens at the transition point with $q_{\rm repel}=q_{\rm repel}^{\rm crit}=1/3$. As the line $\dot{u}_{\rm drift}=0$ and the horizontal line with $\dot{m}_{\rm drift}=0$ touch, the drift close to the fixed point is particularly small allowing for large fluctuations in the steady state. 

The fixed point locations are 
\begin{align}
m_{\rm FP} &=
\begin{cases}
(1-u_{\rm FP}) [\frac{1-2 q_{\rm repel}-3 q_{\rm repel}^2}{(1-q_{\rm repel})^2}]^{1/2} & q_{\rm repel}<\frac{1}{3},\\
0 & {\rm else},
\end{cases}\label{eq:m_FP}\\
u_{\rm FP} &=
\begin{cases}
\frac{q_{\rm doubt} }{q_{\rm doubt}  + (1/q_{\rm repel}-1)}  & q_{\rm repel}<\frac{1}{3},\\
\frac{q_{\rm doubt} }{q_{\rm doubt}  + 2 } & {\rm else}.\end{cases}\label{eq:u_FP}
\end{align}
The phase transition depends on $q_{\rm repel}$ only, $q_{\rm doubt}$ has no influence on the formation of a clear majority opinion. As the drift dominates the model dynamics in the thermodynamic limit $N\to \infty$, we characterize the phase transition by analyzing the dependence of the order parameter $m_{\rm FP}$ on the control parameter $q_{\rm repel}$ in Eq.~(\ref{eq:m_FP}) around the transition point $q_{\rm repel}^{\rm crit}$. In the sub-critical phase $q_{\rm repel}<q_{\rm repel}^{\rm crit}$ close to the phase transition the scaling 
\begin{align}
m_{\rm FP}\propto (q_{\rm repel}^{\rm crit}-q_{\rm repel})^\beta \quad {\rm with} \quad \beta=1/2
\end{align}
holds. Therefore, the phase transition is of continuous mean-field type.

\section{Parameter estimation and further discussion of poll data}

We can use our results for fixed points of the model drift to estimate model parameters from poll data. The repeated poll of Fig.~\ref{fig:poll1} about the EU crisis management of German chancellor Angela Merkel reaches a steady state with clear majority at the beginning of 2012. Using average opinion shares in the steady state as an estimate of $u_{\rm FP}$ and $m_{\rm FP}$, we can solve Eq.~(\ref{eq:m_FP}) for $q_{\rm repel}$ and with this result calculate $q_{\rm doubt}$ with Eq.~(\ref{eq:u_FP}). Averages in the steady state beginning from 2012 are $u=5\%=0.05$ and $m=59\%-36\%=0.23$. The model parameter $q_{\rm repel}=0.327$ is found in the sub-critical phase slightly below the critical value. With small $q_{\rm doubt}=0.109$, the transition of an agent losing a decided opinion is suppressed. 

For the time dependent Brexit poll of Fig.~\ref{fig:poll2} we find $q_{\rm repel}\geq 1/3$, because no clear majority opinion forms. This is in agreement with a highly controversial and emotional debate, where arguments are interpreted individually and therefore can easily repel. The Brexit referendum on 23 June 2016 came out with a very close result of 51.89\% for leave and 48.11\% for remain. This was neither a persistent nor a clear majority decision. The time period before the referendum has an average share of undecided answers of 20.9\%, with Eq.~(\ref{eq:u_FP}) we find $q_{\rm doubt}=0.528$. The time period after the referendum has an average share of undecided answers of 13.3\% and $q_{\rm doubt}=0.306$ accordingly. This means that events where a decided opinion is lost are less likely after the referendum, possibly because the public debate about actual consequences of the Brexit became more pronounced. The fact that there is no clear persistent majority opinion about the Brexit seems to have a strong effect on politics in the UK since the Brexit referendum. No majority for any practical Brexit procedure or for remaining in the EU was found in the House of Commons. A possible reason could be that Members of Parliament have a lot to lose if they decide against half of the electorate. 

Estimating model parameters from poll data without time resolution is harder, as opinion shares can be far from the fixed points. This is the case for Brexit polls in 2012 with a clear majority, convergence to fifty-fifty stalemate happened only until 2014 (Fig.~\ref{fig:poll2}). Deviations can also be due to fluctuations around the fixed point as in the beginning of 2015 for Brexit polls. On the other hand, we found with Figs.~\ref{fig:triangle1} and \ref{fig:triangle2} that the share of undecided agents is closely following the line $\dot{u}_{\rm drift}=0$ even far from stable fixed points. Therefore we can estimate the parameter $q_{\rm doubt}$ even if it is impossible to find $q_{\rm repel}$ from a single poll result without time resolution. If $q_{\rm doubt}$ is changing over time as in Brexit polls, the fraction of undecided agents will always adapt quickly. 
Therefore, we can estimate with Eq.~(\ref{eq:u_dot}) and $\dot{u}_{\rm drift}=0$
\begin{align}
    q_{\rm doubt} &\approx \frac{2u}{1-u-\frac{m^2}{1-u}\,\frac{1-q_{\rm repel}}{1+q_{\rm repel}}}.\label{eq:q_doubt}
\end{align}
Here $u$ and $m$ can be taken directly from the poll data, even if that means values with fluctuations and far from the stable fixed points. 
As the parameter $q_{\rm repel}$ is hard to estimate we use random numbers drawn from a density $\rho(q_{\rm repel})\propto \exp(-q_{\rm repel}/q_0)$ with $0\leq q_{\rm repel}\leq 1$. 

\begin{figure}[htb]
\begin{center}
    \includegraphics[trim=20 0 0 0,clip,width=1.0\columnwidth]{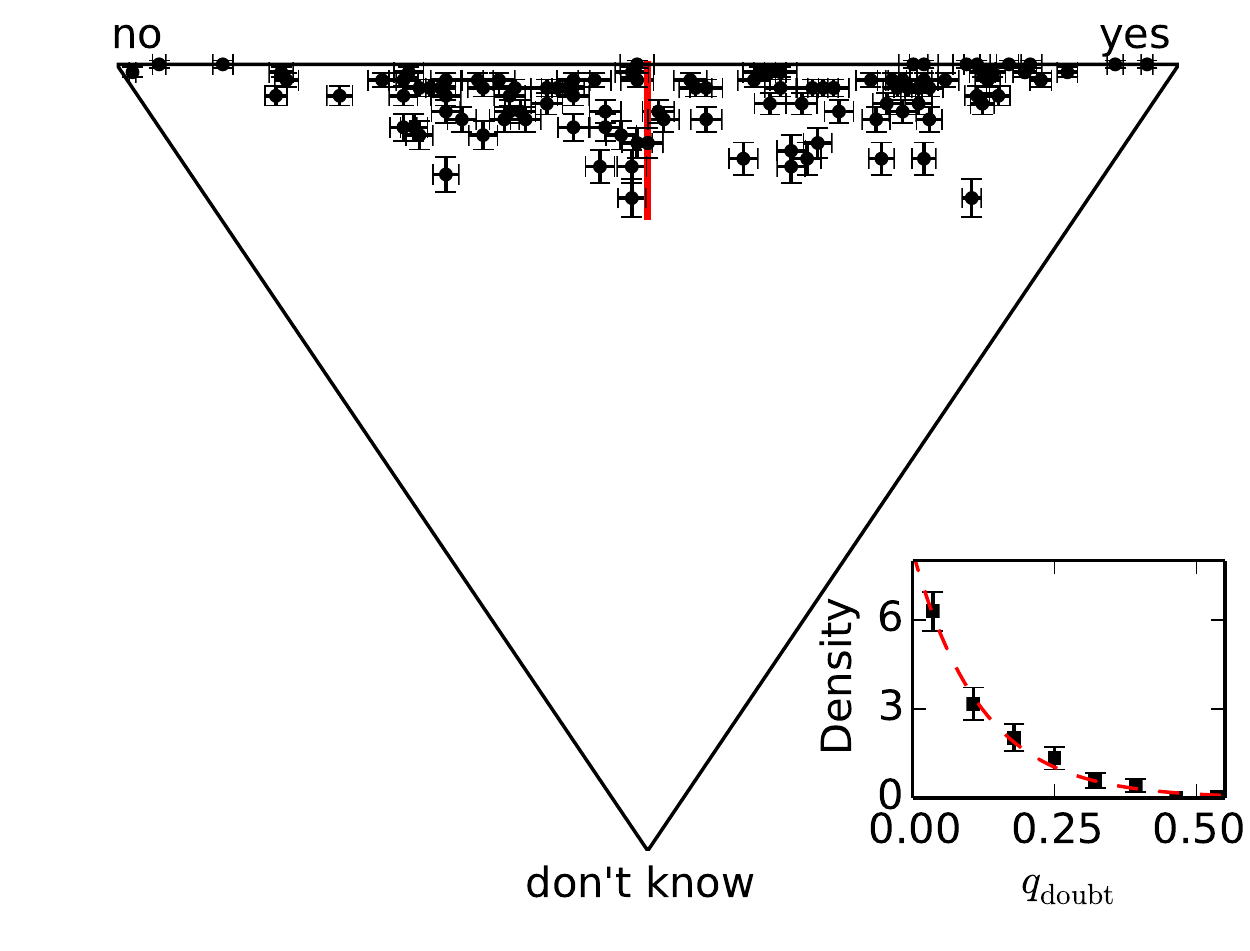}
	\caption{The symbols show results of all the polls performed by ``Infratest Dimap'' \cite{infratest} in the years 2013 and 2014, with possible answers ``yes'', ``no'' or ``don't know'' (112 polls). Shown are the respective frequencies, errorbars are two standard deviations. 
	The red line shows the parameter line for states without majority opinion, $q_{\rm repel}\geq 1/3$. 
    In the inset on the lower right, a binning of the density of the model parameter $q_{\rm doubt}$ is shown with black circles.  
    The density is compatible with an exponential density $\propto \exp(-q_{\rm doubt}/0.12)$ (red dashed line).}	\label{fig:polldata}
\end{center}
\end{figure}

The poll data in Fig.~\ref{fig:polldata} have a broad range of majority opinion shares, together with small but significant fractions of undecided respondents. The density of parameters $q_{\rm doubt}$ is shown with the inset on the lower right of Figure~\ref{fig:polldata} for $q_0=0.1$ and error bars of one standard deviation with the jackknife method. We tested that this result changes only slightly with $q_0$ by using $q_0=0.01; 0.1; 1$. The density of $q_{\rm doubt}$ is compatible with an exponential density with typical parameter of $q_{\rm doubt}=0.12\pm 0.03$. This means for our model that doubt in the presence of an opposite opinion is a significant effect. However, it is suppressed compared to convincing events for undecided agents. 

For the parameter $q_{\rm repel}$, we can estimate a plausible region $0\leq q_{\rm repel} \leq 1$. With Eq. \ref{eq:u_FP} we see that fixed points are sensitive to this parameter only for $q_{\rm repel} < \frac{1}{3}$. The reasoning for the upper bound is as follows: Our model definition is plausible only for $q_{\rm repel}\leq 1$, else discussions would more often repel than convince an undecided agent. Many polls have a clear majority opinion, indicating that $q_{\rm repel}<1/3$ is common. 
On the other hand, eight out of 112 poll results are compatible with fifty-fifty states implying $q_{\rm repel}\geq 1/3$ (the line of such fixed points, indicated with the red line, is within two standard deviations of the result).
Therefore, $q_{\rm repel}$ in our model has to take surprisingly high values in order to reproduce these polls: In more then one out of four cases, a discussion would lead to repelling an undecided agent instead of convincing. 
Examples for questions with no clear majority opinion are: "Is the worldwide political situation threatening for people in Germany?" and "Immigration to Germany should be restricted". 
Two more referendums in Europe which were important and controversial also showed results without clear majority: The Scottish independence referendum in 2014 ended with 44.7\% voting for independence and a majority against it. 
The Swiss referendum "against mass immigration" in February 2014 had 50.3\% agreement. 

Questions with the clearest consensus are "Help for poor families: Free books in schools" (97\% yes) and "It would be a problem, if in my backyard there was a retirement home" (98\% no). 
According to the share of undecided agents, the parameter $q_{\rm doubt}$ might also indicate the lack of interest towards a topic. For example, the question "Are canonizations of the catholic church contemporary?" has 14\% undecided answers (24\% yes), while the statement "I am worried about my savings" is intuitively highly relevant for most respondents, and results in only 2\% undecided answers (48\% yes).

\section{Conclusion}

We analyzed a voter model variant, where an additional undecided state of agents 
opinions was introduced. The outcome of discussions among pairs of agents was given 
in a way allowing for repelling and doubt effects. The model dynamics in the 
presence of strong repulsion lead to a fifty-fifty stalemate where no opinion could 
win in the long run. As we expect repulsion effects to be particularly strong in 
controversial discussions, this means that in controversial debates it is hard to 
find persistent clear majorities making it hard for politics to find a broadly 
accepted consensus. Based on repeated poll data about the Brexit over more than seven 
years we were able to identify such a trapped fifty-fifty stalemate. The poll 
data also included shares of undecided respondents which were surprisingly constant 
over time but dropped fast around the actual Brexit referendum. This finding agreed 
with our model dynamics having a time scale separation, and it was in line 
with the interpretation of our model parameters. After the Brexit referendum it 
was less likely to lose a decided opinion because implications of the Brexit 
became more clear and real. 
Also comparing to further poll data we found that opinion switches away from another 
opinion, as well as switches due to doubt can be surprisingly likely. 

In our model we ignored transitions between extremes. This restriction is not substantial, 
because in an extended model including convincing transitions directly from 
``yes'' to ``no'' or the other way around, the dynamics of opinion shares is almost the same. 
It would be interesting to study further model modifications as individual model parameters for every agent, social interaction networks or broadcasting effects. This could result in increased fluctuations even for large populations. 

The model dynamics observed here has serious implications for the understanding 
of controversial debates. First of all, making debates even more controversial will 
not help to find a clear majority but will keep both opinions strong. Second, making
a debate more detailed and basing it on facts can help finding a clear majority. If 
this is not enough to escape from the fifty-fifty stalemate, at least it becomes easier 
to find possible compromises for politics which are acceptable for a broad majority. 
This can be enabled by making referendum decisions more detailed. An 
example is the referendum in Switzerland against mass immigration of 2014 which ended 
close to parity but still the consequences were handled well by politics.

\begin{acknowledgments}
The authors would like to thank Jan Lorenz for helpful comments.  
\end{acknowledgments}

\bibliography{bibli}

\begin{thebibliography}{25}
\expandafter\ifx\csname natexlab\endcsname\relax\def\natexlab#1{#1}\fi
\expandafter\ifx\csname bibnamefont\endcsname\relax
  \def\bibnamefont#1{#1}\fi
\expandafter\ifx\csname bibfnamefont\endcsname\relax
  \def\bibfnamefont#1{#1}\fi
\expandafter\ifx\csname citenamefont\endcsname\relax
  \def\citenamefont#1{#1}\fi
\expandafter\ifx\csname url\endcsname\relax
  \def\url#1{\texttt{#1}}\fi
\expandafter\ifx\csname urlprefix\endcsname\relax\def\urlprefix{URL }\fi
\providecommand{\bibinfo}[2]{#2}
\providecommand{\eprint}[2][]{\url{#2}}

\bibitem[{\citenamefont{Page et~al.}(1987)\citenamefont{Page, Shapiro, and
  Dempsey}}]{page1987moves}
\bibinfo{author}{\bibfnamefont{B.~I.} \bibnamefont{Page}},
  \bibinfo{author}{\bibfnamefont{R.~Y.} \bibnamefont{Shapiro}},
  \bibnamefont{and} \bibinfo{author}{\bibfnamefont{G.~R.}
  \bibnamefont{Dempsey}}, \bibinfo{journal}{American Political Science Review}
  \textbf{\bibinfo{volume}{81}}, \bibinfo{pages}{23} (\bibinfo{year}{1987}).

\bibitem[{\citenamefont{Barber{\'a} et~al.}(2015)\citenamefont{Barber{\'a},
  Jost, Nagler, Tucker, and Bonneau}}]{barbera2015tweeting}
\bibinfo{author}{\bibfnamefont{P.}~\bibnamefont{Barber{\'a}}},
  \bibinfo{author}{\bibfnamefont{J.~T.} \bibnamefont{Jost}},
  \bibinfo{author}{\bibfnamefont{J.}~\bibnamefont{Nagler}},
  \bibinfo{author}{\bibfnamefont{J.~A.} \bibnamefont{Tucker}},
  \bibnamefont{and} \bibinfo{author}{\bibfnamefont{R.}~\bibnamefont{Bonneau}},
  \bibinfo{journal}{Psychological science} \textbf{\bibinfo{volume}{26}},
  \bibinfo{pages}{1531} (\bibinfo{year}{2015}).

\bibitem[{\citenamefont{Del~Vicario et~al.}(2016)\citenamefont{Del~Vicario,
  Bessi, Zollo, Petroni, Scala, Caldarelli, Stanley, and
  Quattrociocchi}}]{del2016spreading}
\bibinfo{author}{\bibfnamefont{M.}~\bibnamefont{Del~Vicario}},
  \bibinfo{author}{\bibfnamefont{A.}~\bibnamefont{Bessi}},
  \bibinfo{author}{\bibfnamefont{F.}~\bibnamefont{Zollo}},
  \bibinfo{author}{\bibfnamefont{F.}~\bibnamefont{Petroni}},
  \bibinfo{author}{\bibfnamefont{A.}~\bibnamefont{Scala}},
  \bibinfo{author}{\bibfnamefont{G.}~\bibnamefont{Caldarelli}},
  \bibinfo{author}{\bibfnamefont{H.~E.} \bibnamefont{Stanley}},
  \bibnamefont{and}
  \bibinfo{author}{\bibfnamefont{W.}~\bibnamefont{Quattrociocchi}},
  \bibinfo{journal}{Proceedings of the National Academy of Sciences}
  \textbf{\bibinfo{volume}{113}}, \bibinfo{pages}{554} (\bibinfo{year}{2016}).

\bibitem[{\citenamefont{Castellano et~al.}(2009)\citenamefont{Castellano,
  Fortunato, and Loreto}}]{castellano2009statistical}
\bibinfo{author}{\bibfnamefont{C.}~\bibnamefont{Castellano}},
  \bibinfo{author}{\bibfnamefont{S.}~\bibnamefont{Fortunato}},
  \bibnamefont{and} \bibinfo{author}{\bibfnamefont{V.}~\bibnamefont{Loreto}},
  \bibinfo{journal}{Reviews of modern physics} \textbf{\bibinfo{volume}{81}},
  \bibinfo{pages}{591} (\bibinfo{year}{2009}).

\bibitem[{\citenamefont{Clifford and Sudbury}(1973)}]{clifford_model_1973}
\bibinfo{author}{\bibfnamefont{P.}~\bibnamefont{Clifford}} \bibnamefont{and}
  \bibinfo{author}{\bibfnamefont{A.}~\bibnamefont{Sudbury}},
  \bibinfo{journal}{Biometrika} \textbf{\bibinfo{volume}{60}},
  \bibinfo{pages}{581} (\bibinfo{year}{1973}), ISSN \bibinfo{issn}{0006-3444,
  1464-3510},
  \urlprefix\url{http://biomet.oxfordjournals.org/content/60/3/581}.

\bibitem[{\citenamefont{Holley and Liggett}(1975)}]{holley_ergodic_1975}
\bibinfo{author}{\bibfnamefont{R.~A.} \bibnamefont{Holley}} \bibnamefont{and}
  \bibinfo{author}{\bibfnamefont{T.~M.} \bibnamefont{Liggett}},
  \bibinfo{journal}{The Annals of Probability} \textbf{\bibinfo{volume}{3}},
  \bibinfo{pages}{643} (\bibinfo{year}{1975}), ISSN \bibinfo{issn}{0091-1798},
  \bibinfo{note}{{ArticleType:} research-article / Full publication date: Aug.,
  1975 / Copyright © 1975 Institute of Mathematical Statistics},
  \urlprefix\url{http://www.jstor.org/stable/2959329}.

\bibitem[{\citenamefont{Redner}(2001)}]{redner2001guide}
\bibinfo{author}{\bibfnamefont{S.}~\bibnamefont{Redner}},
  \emph{\bibinfo{title}{A guide to first-passage processes}}
  (\bibinfo{publisher}{Cambridge University Press}, \bibinfo{year}{2001}).

\bibitem[{\citenamefont{Dornic et~al.}(2001)\citenamefont{Dornic, Chaté,
  Chave, and Hinrichsen}}]{dornic_critical_2001}
\bibinfo{author}{\bibfnamefont{I.}~\bibnamefont{Dornic}},
  \bibinfo{author}{\bibfnamefont{H.}~\bibnamefont{Chaté}},
  \bibinfo{author}{\bibfnamefont{J.}~\bibnamefont{Chave}}, \bibnamefont{and}
  \bibinfo{author}{\bibfnamefont{H.}~\bibnamefont{Hinrichsen}},
  \bibinfo{journal}{Physical Review Letters} \textbf{\bibinfo{volume}{87}},
  \bibinfo{pages}{045701} (\bibinfo{year}{2001}),
  \urlprefix\url{http://link.aps.org/doi/10.1103/PhysRevLett.87.045701}.

\bibitem[{\citenamefont{Krause et~al.}(2012)\citenamefont{Krause, B\"{o}ttcher,
  and Bornholdt}}]{krause_mean-field-like_2012}
\bibinfo{author}{\bibfnamefont{S.~M.} \bibnamefont{Krause}},
  \bibinfo{author}{\bibfnamefont{P.}~\bibnamefont{B\"{o}ttcher}},
  \bibnamefont{and}
  \bibinfo{author}{\bibfnamefont{S.}~\bibnamefont{Bornholdt}},
  \bibinfo{journal}{Physical Review E} \textbf{\bibinfo{volume}{85}},
  \bibinfo{pages}{031126} (\bibinfo{year}{2012}),
  \urlprefix\url{http://link.aps.org/doi/10.1103/PhysRevE.85.031126}.

\bibitem[{\citenamefont{Holme and Newman}(2006)}]{holme2006nonequilibrium}
\bibinfo{author}{\bibfnamefont{P.}~\bibnamefont{Holme}} \bibnamefont{and}
  \bibinfo{author}{\bibfnamefont{M.~E.} \bibnamefont{Newman}},
  \bibinfo{journal}{Physical Review E} \textbf{\bibinfo{volume}{74}},
  \bibinfo{pages}{056108} (\bibinfo{year}{2006}).

\bibitem[{\citenamefont{Krause and Bornholdt}(2012)}]{krause_opinion_2012}
\bibinfo{author}{\bibfnamefont{S.~M.} \bibnamefont{Krause}} \bibnamefont{and}
  \bibinfo{author}{\bibfnamefont{S.}~\bibnamefont{Bornholdt}},
  \bibinfo{journal}{Physical Review E} \textbf{\bibinfo{volume}{86}},
  \bibinfo{pages}{056106} (\bibinfo{year}{2012}),
  \urlprefix\url{http://link.aps.org/doi/10.1103/PhysRevE.86.056106}.

\bibitem[{\citenamefont{Araujo et~al.}(2010)\citenamefont{Araujo, Andrade~Jr,
  and Herrmann}}]{araujo2010tactical}
\bibinfo{author}{\bibfnamefont{N.~A.} \bibnamefont{Araujo}},
  \bibinfo{author}{\bibfnamefont{J.~S.} \bibnamefont{Andrade~Jr}},
  \bibnamefont{and} \bibinfo{author}{\bibfnamefont{H.~J.}
  \bibnamefont{Herrmann}}, \bibinfo{journal}{PLoS One}
  \textbf{\bibinfo{volume}{5}}, \bibinfo{pages}{e12446} (\bibinfo{year}{2010}).

\bibitem[{\citenamefont{Fern{\'a}ndez-Gracia
  et~al.}(2014)\citenamefont{Fern{\'a}ndez-Gracia, Suchecki, Ramasco,
  San~Miguel, and Egu{\'\i}luz}}]{fernandez2014voter}
\bibinfo{author}{\bibfnamefont{J.}~\bibnamefont{Fern{\'a}ndez-Gracia}},
  \bibinfo{author}{\bibfnamefont{K.}~\bibnamefont{Suchecki}},
  \bibinfo{author}{\bibfnamefont{J.~J.} \bibnamefont{Ramasco}},
  \bibinfo{author}{\bibfnamefont{M.}~\bibnamefont{San~Miguel}},
  \bibnamefont{and} \bibinfo{author}{\bibfnamefont{V.~M.}
  \bibnamefont{Egu{\'\i}luz}}, \bibinfo{journal}{Physical review letters}
  \textbf{\bibinfo{volume}{112}}, \bibinfo{pages}{158701}
  (\bibinfo{year}{2014}).

\bibitem[{\citenamefont{Vazquez and Redner}(2004)}]{vazquez2004ultimate}
\bibinfo{author}{\bibfnamefont{F.}~\bibnamefont{Vazquez}} \bibnamefont{and}
  \bibinfo{author}{\bibfnamefont{S.}~\bibnamefont{Redner}},
  \bibinfo{journal}{Journal of Physics A: Mathematical and General}
  \textbf{\bibinfo{volume}{37}}, \bibinfo{pages}{8479} (\bibinfo{year}{2004}).

\bibitem[{\citenamefont{Vazquez et~al.}(2003)\citenamefont{Vazquez, Krapivsky,
  and Redner}}]{vazquez2003constrained}
\bibinfo{author}{\bibfnamefont{F.}~\bibnamefont{Vazquez}},
  \bibinfo{author}{\bibfnamefont{P.~L.} \bibnamefont{Krapivsky}},
  \bibnamefont{and} \bibinfo{author}{\bibfnamefont{S.}~\bibnamefont{Redner}},
  \bibinfo{journal}{Journal of Physics A: Mathematical and General}
  \textbf{\bibinfo{volume}{36}}, \bibinfo{pages}{L61} (\bibinfo{year}{2003}).

\bibitem[{\citenamefont{Castell{\'o} et~al.}(2006)\citenamefont{Castell{\'o},
  Egu{\'\i}luz, and San~Miguel}}]{castello2006ordering}
\bibinfo{author}{\bibfnamefont{X.}~\bibnamefont{Castell{\'o}}},
  \bibinfo{author}{\bibfnamefont{V.~M.} \bibnamefont{Egu{\'\i}luz}},
  \bibnamefont{and}
  \bibinfo{author}{\bibfnamefont{M.}~\bibnamefont{San~Miguel}},
  \bibinfo{journal}{New Journal of Physics} \textbf{\bibinfo{volume}{8}},
  \bibinfo{pages}{308} (\bibinfo{year}{2006}).

\bibitem[{\citenamefont{Castell{\'o} et~al.}(2009)\citenamefont{Castell{\'o},
  Baronchelli, and Loreto}}]{castello2009consensus}
\bibinfo{author}{\bibfnamefont{X.}~\bibnamefont{Castell{\'o}}},
  \bibinfo{author}{\bibfnamefont{A.}~\bibnamefont{Baronchelli}},
  \bibnamefont{and} \bibinfo{author}{\bibfnamefont{V.}~\bibnamefont{Loreto}},
  \bibinfo{journal}{The European Physical Journal B}
  \textbf{\bibinfo{volume}{71}}, \bibinfo{pages}{557} (\bibinfo{year}{2009}).

\bibitem[{\citenamefont{Dall'Asta and Castellano}(2007)}]{dall2007effective}
\bibinfo{author}{\bibfnamefont{L.}~\bibnamefont{Dall'Asta}} \bibnamefont{and}
  \bibinfo{author}{\bibfnamefont{C.}~\bibnamefont{Castellano}},
  \bibinfo{journal}{EPL (Europhysics Letters)} \textbf{\bibinfo{volume}{77}},
  \bibinfo{pages}{60005} (\bibinfo{year}{2007}).

\bibitem[{\citenamefont{Dall'Asta and Galla}(2008)}]{dall2008algebraic}
\bibinfo{author}{\bibfnamefont{L.}~\bibnamefont{Dall'Asta}} \bibnamefont{and}
  \bibinfo{author}{\bibfnamefont{T.}~\bibnamefont{Galla}},
  \bibinfo{journal}{Journal of Physics A: Mathematical and Theoretical}
  \textbf{\bibinfo{volume}{41}}, \bibinfo{pages}{435003}
  (\bibinfo{year}{2008}).

\bibitem[{\citenamefont{Malarz and Kulakowski}(2009)}]{malarz2009indifferents}
\bibinfo{author}{\bibfnamefont{K.}~\bibnamefont{Malarz}} \bibnamefont{and}
  \bibinfo{author}{\bibfnamefont{K.}~\bibnamefont{Kulakowski}},
  \bibinfo{journal}{arXiv preprint arXiv:0908.3387}  (\bibinfo{year}{2009}).

\bibitem[{\citenamefont{Svenkeson and Swami}(2015)}]{Svenkeson2015reaching}
\bibinfo{author}{\bibfnamefont{A.}~\bibnamefont{Svenkeson}} \bibnamefont{and}
  \bibinfo{author}{\bibfnamefont{A.}~\bibnamefont{Swami}},
  \bibinfo{journal}{Submitted}  (\bibinfo{year}{2015}).

\bibitem[{\citenamefont{Vasilopoulou}(2016)}]{vasilopoulou2016uk}
\bibinfo{author}{\bibfnamefont{S.}~\bibnamefont{Vasilopoulou}},
  \bibinfo{journal}{The Political Quarterly} \textbf{\bibinfo{volume}{87}},
  \bibinfo{pages}{219} (\bibinfo{year}{2016}).

\bibitem[{\citenamefont{Galam}(2004)}]{galam2004contrarian}
\bibinfo{author}{\bibfnamefont{S.}~\bibnamefont{Galam}},
  \bibinfo{journal}{Physica A: Statistical Mechanics and its Applications}
  \textbf{\bibinfo{volume}{333}}, \bibinfo{pages}{453} (\bibinfo{year}{2004}).

\bibitem[{inf()}]{infratest}
\emph{\bibinfo{title}{Yougov public limited company}},
  \bibinfo{howpublished}{\url
  {http://www.infratest-dimap.de/umfragen-analysen/bundesweit/umfragen/}}.

\bibitem[{bre()}]{brexit}
\emph{\bibinfo{title}{Infratest dimap gesellschaft f\"{u}r trend- und
  wahlforschung mbh}}, \bibinfo{howpublished}{\url
  {https://whatukthinks.org/eu/questions}}.

\end{thebibliography}

%
%

\end{document}